# The Application of Affective Measures in Text-based Emotion Aware Recommender Systems

John Kalung Leung[1][a], Igor Griva[2][b], William G. Kennedy[3][c], Jason M. Kinser[4][d],
Sohyun Park[1][e], and Seo Young Lee[5][f]

[1]*Computational Sciences and Informatics, Computational and Data Sciences Department, George Mason University Korea, 119-4 Songdomunhwa-ro, Yeonsu-gu, Incheon, 21985, Korea*

[2]*Department of Mathematical Sciences, MS3F2, Exploratory Hall 4114, George Mason University, 4400 University Drive, Fairfax, Virginia 22030, USA*

[3]*Center for Social Complexity, Computational and Data Sciences Department, College of Science, George Mason University, 4400 University Drive, Fairfax, Virginia 22030, USA*

[4]*Computational Sciences and Informatics, Computational and Data Sciences Department, College of Science, George Mason University, 4400 University Drive, Fairfax, Virginia 22030, USA*

[5]*Department of Communications, George Mason University Korea, 119-4 Songdomunhwa-ro, Yeonsu-gu, Incheon, 21985, Korea*

{jleung2, igriva, wkennedy, jkinser, spark230, slee336}@gmu.edu

Abstract: This paper presents an innovative approach to address the problems researchers face in Emotion Aware Recommender Systems (EARS): the difficulty and cumbersome collecting voluminously good quality emotion-tagged datasets and an effective way to protect users' emotional data privacy. Without enough good-quality emotion-tagged datasets, researchers cannot conduct repeatable affective computing research in EARS that generates personalized recommendations based on users' emotional preferences. Similarly, if we fail to protect users' emotional data privacy fully, users could resist engaging with EARS services. This paper introduced a method that detects affective features in subjective passages using the Generative Pre-trained Transformer Technology, forming the basis of the Affective Index and Affective Index Indicator (AII). Eliminate the need for users to build an affective feature detection mechanism. The paper advocates for a separation of responsibility approach where users protect their emotional profile data while EARS service providers refrain from retaining or storing it. Service providers can update users' Affective Indices in memory without saving their privacy data, providing Affective Aware recommendations without compromising user privacy. This paper offers a solution to the subjectivity and variability of emotions, data privacy concerns, and evaluation metrics and benchmarks, paving the way for future EARS research.

## 1 INTRODUCTION

The Emotion Aware Recommender System (EARS) provides personalized recommendations based on users' affective preferences and opinions of other similar users. However, EARS research faces various challenges due to human emotions' inherent subjectivity and variability (Qian et al., 2019). Difficulties include developing accurate emotion detection, classification, and prediction models and collecting sufficient emotion-tagged datasets due to emotions' subjective and contextual nature (Schedl et al., 2018). Moreover, balancing personalized recommendations with ethical considerations around user privacy and data protection is another vital challenge in EARS research (Bobadilla et al., 2013). Robust evaluation metrics and benchmarks are necessary to assess the effectiveness of EARS, as defining and measuring the impact of emotions on user behavior and satisfaction is complex and multifaceted (Ghazi et al., 2015), and (Mohammad, 2016). This paper devises an innovative approach to address the following problems researchers face in EARS: the difficulty and cumbersome collecting voluminously good quality emotion-tagged datasets and an effective way to protect users' emotional data privacy.

[a]  https://orcid.org/0000-0003-0216-1134
[b]  https://orcid.org/0000-0002-2291-233X
[c]  https://orcid.org/0000-0001-9238-1215
[d]  https://orcid.org/0000-0003-0078-4899
[e]  https://orcid.org/0000-0002-1231-5662
[f]  https://orcid.org/ 0000-0001-5867-0726

## 2 AFFECTIVE TAGGED DATASETS

EARS needs voluminously good quality emotional tags datasets for model training of making personalized recommendations. EARS requires Emotional tags to refer to labeling data, such as Ekman's six basic emotions: happiness, sadness, anger, fear, surprise, and disgust (Humintell, 2020). These tags are crucial in developing EARS technology that can provision users' and items' emotional profiles. However, collecting accurate and sufficient emotional tags can be extremely challenging (Scherer, 2005) due to subjective and variable emotions (Russell, 2003), lack of standard tagging methods (Borth et al., 2013), time and resource intensity (Lo et al., 2017), privacy issues (Kompan et al., 2015), limited diversity (Yang et al., 2021), and contextual factors (Kanjo et al., 2015). Therefore, collecting emotional tags for EARS research demands thoughtful planning and consideration of these factors (Mauss & Robinson, 2009b) in data features engineering.

## 3. AFFECTIVE INDEX AND AFFECTIVE INDEX INDICATOR

The Affective Index is used in Recommender Systems to model the emotional profiles of users and items. It measures the human emotion metrics in terms of probabilistic values (Acheampong et al., n.d.). The Affective Index Indicator (AII), however, advocated by Leung et al. (Leung et al., 2021a), generates a list of numerical similarity values by comparing the affective attributes of a source object against a list of target objects using the Cosine Similarity metrics (Leung et al., 2020b) or Nearest Neighbor algorithm (Keller et al., 1985). AII approaches have evolved from simple word lists (Kratzwald et al., 2018) to more sophisticated machine learning (Nasir et al., 2020) and lexicon-based methods (Hajek et al., 2020), and some recent research combines multiple techniques to create a more robust AII (Naseem et al., 2020). However, challenges in using AII include reliability, interpretability, and ethical concerns around privacy, stereotypes, and biases (Mauss & Robinson, 2009a). The ongoing development of more accurate and effective methods for measuring emotions in the text (Leung et al., 2020a) can enhance the relevancy and accuracy of personalized EARS.

## 4. APPLYING AFFECTIVE INDEX INDICATOR IN EARS

Affective Index Indicator (AII) is a metric used in Emotion Aware Recommender Systems (EARS) to measure the emotional content of textual data. It reflects the intensity of emotion words expressed in a text, such as happiness, sadness, anger, fear, surprise, and disgust. Various natural language processing techniques analyze the sentiment and emotion in a text to calculate the AII (Tsytsarau & Palpanas, 2012). By considering the emotional preferences of users and items, recommender systems can provide personalized recommendations that better-fit users' needs and preferences (Chang & Hsing, 2021). AII is valuable for building EARS because it allows the system to consider the emotional profile of items and an active user's affective preferences in making recommendations. However, AII is one of many approaches to building EARS, and its effectiveness may vary depending on the specific application and user context. When designing EARS, designers must consider ethical and privacy considerations to ensure the system does not perpetuate biases or stereotypes related to emotions or personal characteristics and to secure user data.

## 5. THE BIGGEST CHALLENGES IN EARS RESEARCH

The main challenge in text-based Emotion Aware Recommender Systems (EARS) research is the difficulty in obtaining high-quality, emotion-tagged datasets, which are necessary for machine learning processing. To address this challenge, Guo (Guo, 2022) illustrated a deep learning-assisted semantic text analysis approach that involves defining emotion keywords, identifying data sources, developing a collection plan, cleaning and preprocessing data, and evaluating and refining the dataset. However, researchers still face a shortage of high-quality emotion-labeled datasets, which are needed to train the emotion prediction model for classification. While standards Recent advances in transformer-based models, such as Generative Pre-trained Transformer (GPT) technology, have yet to establish benchmarking datasets for generating, PaLM (Chowdhery et al., 2022), GPT-3 (Brown et al., 2020), ChatGPT (Zhou, Li, Li, Yu, Liu, Wang, Zhang, Ji, Yan, He, et al., 2023), BERT (Devlin et al., 2018), ELMO (Grossglauser et al., 2018), RoBERTa (Liu et al., 2019), and Transformer-XL (Nassiri & Akhloufi, 2022) offer promising new approaches to dataset collection (Ethayarajh, n.d.). These models have been trained on large amounts of text data to generate emotion labels (Kusal, Patil, Choudrie, Kotecha, Mishra, et al., 2022). They may provide valuable resources for researchers in this field.

Instead of devising a method to collect the needed emotion-tagged dataset for EARS research, this paper advocates leveraging the GPT extensive database stored with multiple domains of information and query through conversational dialogue for the Affective Index of a subjective text without extra coding effort.

## 6. GATHERING AFFECTIVE INDICES AND BUILDING AFFECTIVE INDEX INDICATORS USING GPT TECHNOLOGY

Instead of devising a method to collect the needed emotion-tagged dataset for EARS research, this paper advocates leveraging the GPT extensive database stored with multiple domains of information and query through conversational dialogue for the Affective Index of a subjective text.

Sometimes, the GPT database may have stored the domain information researchers seek. For example, GPT held an extensive collection of movie plots which makes the gathering effort of the Affective Indices for these movie plots effortlessly. Once GPT confirms it has stored the movie plot, a researcher may ask GPT for the associate Affective Index of the story. Otherwise, research can supply the movie plot to GPT for predicting the associate Affective Index. Using GPT conversational dialogue to extract Affective indices from an object descriptive passage not only proves effortlessly, but it may also, if embraced by many, become the de facto standard way to gather an emotional label for any object through its subjective description text.

Recommender Systems use the terms Affective Index and Affective Index Indicator (AII) to measure the emotions of users and items by measuring probabilities of Ekman's six basic human emotions like happiness or anger (Leung et al., 2020b). The AII method creates a list of numbers that indicates how much a source object is like a group of target objects with similar features, using the Cosine Similarity or Nearest Neighbour algorithm. AII can measure emotions in text using various techniques such as word lists or machine learning algorithms, but using it poses challenges, such as ensuring the indicators are reliable and easy to understand and avoiding reinforcing stereotypes or biases. Further research is necessary to enhance AII. Affective computing has gained significant attention in recent years, with researchers aiming to develop standardized methods for gathering affective indices from subjective passages. In this regard, artificial intelligence language models like ChatGPT have shown promising results in sentiment and affective analysis of a text ( Kocoń et al., 2023). This study illustrates the use of short prompting to query ChatGPT to build an Affective Index from the inspirational text (Thorp, 2023). This research addresses these challenges and develops better ways to measure emotions in text.

Firstly, the subjective passages must undergo preprocessing to ensure a suitable format for sentiment analysis. This process involves removing irrelevant information or noise, such as stop words or punctuation, and tokenizing the text into individual words or phrases. Subsequently, ChatGPT can analyze the sentiment of each word or phrase in the subjective passages by assigning a polarity score, indicating whether it expresses a positive or negative opinion ( Kocoń et al., 2023). Several libraries and tools are available for sentiment analysis, including NLTK (Yao, 2019), TextBlob (Gujjar & Kumar, 2021), and VADER (Song et al., 2020).

Once you obtain the polarity scores for each word or phrase, you can aggregate them to calculate an overall AII for the subjective passages. One common approach is to take the average polarity score across all words or phrases. Various cross-domain downstream applications can use the AII (Leung et al., 2021b), such as filtering out subjective passages likely to be poorly received by users or providing personalized recommendations matching users' emotional preferences.

ChatGPT can estimate the probabilistic values of basic emotions present in a subjective passage, such as Ekman's basic six human emotions, using sentiment analysis techniques (Munn et al., 2023). To do so, ChatGPT tokenizes and preprocesses the text and utilizes pre-existing lexicons or machine learning models to identify and classify the intensity of each emotion. Finally, ChatGPT normalizes the score's probabilistic value of the intensity of each emotion present to create an Affective Index (Lauriola et al., 2022).

While sentiment analysis is not a perfect science, our method provides a helpful guide for assessing emotional content in subjective passages. The AII and Affective Index can be valuable tools for various applications, including recommendation systems and emotional filtering.

# 7. USING CHATGPT TO ESTIMATE THE PROBABILISTIC VALUES FOR THE AFFECTIVE INDEX

We demonstrate the use of ChatGPT to determine the probability of detecting primary human emotions in a given text. Using the "Godfather I" plot as a case study, we asked ChatGPT to compute Ekman's basic human emotions' intensity. It returned the emotion_probs dictionary containing the Affective Index as follows: {'happiness': 0.02571, 'sadness': 0.81373, 'anger': 0.05563, 'fear': 0.09933, 'surprise': 0.00486, 'disgust': 0.00074}. The Affective Index expresses the probabilistic values of detected emotions. We use OpenAI's GPT-3 API to analyze the plot's emotions and extract the scores for each primary emotion as suggested (Dale, 2021). We asked GPT-3 to normalize the scores and compute the probabilistic values of Ekman's six basic human emotions scores. Our example shows a high intensity of sadness and moderate intensity of fear and anger in the movie plot. The Affective Index may vary based on the model and other factors.

# 8. ALTERNATIVE TO GPT TECHNOLOGIES

Several organizations have developed language models and conversational agents with similar purposes. For example, Google's Meena (Thoppilan et al., 2022) uses a transformer-based neural network to generate human-like responses, while Microsoft's XiaoIce ("The Design and Implementation of Xiaoice, an Empathetic Social Chatbot," 2020) emulates a teenage girl's conversational style and personality. Facebook's Blender (Kusal, Patil, Choudrie, Kotecha, Vora, et al., 2022) combines rule-based and machine-learning approaches to generate natural language responses, while Amazon's Alexa (Hoy, 2018) and Apple's Siri (Kepuska & Bohouta, 2018) use natural language processing and machine learning to understand and respond to user requests. Each of these conversational agents has its strengths and weaknesses, and depending on the specific use case, one may be more effective than another.

Other comparable tools include OpenAI's GPT-3 (Maddigan & Susnjak, 2023), which is larger and more potent than ChatGPT ((Wahde & Virgolin, 2022)), Google's BERT (Devlin et al., 2018), which excels at understanding the context of words and phrases, Allen Institute for Artificial Intelligence's ELMO (Grossglauser et al., 2018), which generates contextualized word embeddings, Facebook's RoBERTa (Liu et al., 2019), which is pre-trained on a larger dataset, and Carnegie Mellon University and Google's Transformer-XL (Nassiri & Akhloufi, 2022), which handles longer text sequences and generates more accurate predictions for text completion tasks. Although all of these language models have been trained on large amounts of text data and use variations of the transformer architecture, each model has its strengths and weaknesses and may suit specific tasks or use cases better (Zhou, Li, Li, Yu, Liu, Wang, Zhang, Ji, Yan, & He, 2023).

# 9. APPLICATION OF AFFECTIVE INDEX AND AFFECTIVE INDEX INDICATOR

This study introduces a new method for detecting affective features in subjective writing using a Generative Pre-trained Transformer (GPT) Natural Language Processing (NLP) database. The Affective Index method relies on GPT to create a profile of an item's affective features. To derive the user's affective profile, one can average the Affective Indices of all the consumed objects (Leung et al., 2020c). The Affective Index plays a crucial role in Recommender Systems, where subjective descriptions are utilized, as it can determine the similarity between items and users.

The method involves separating the responsibility for creating and maintaining emotional profiles to protect users' privacy. Users can provide their "emotion ID" to the EARS service provider, who will generate an Affective Index Indicator list based on their emotional profile. The service provider does not store users' affective profile data but instead performs the processing in memory.

While the method has potential applications in various recommendation scenarios, including collaborative-based and content-based recommendations, data privacy is a significant concern. One possible solution is to have users assume responsibility for protecting their emotional profile data. However, further research is needed to address these concerns fully.

## 10. CONCLUSION

The field of Emotion Aware Recommender Systems (EARS) research is a challenging and complex area that aims to provide personalized recommendations based on users' affective preferences. However, this research faces numerous obstacles that hinder its development, including subjectivity and variability of emotions, the absence of established standardized affective tagging schemes, significant time and resource requirements, privacy concerns, limited diversity, and contextual factors. These challenges make it difficult to gather affective tagged datasets for EARS research.

This study highlighted the use of a Generative Pre-trained Transformer NLP database, specifically ChatGPT, to detect affective features and generate Affective Indices. The Affective Index Indicator (AII) method assesses human emotions expressed as probability values. This method has undergone various advancements, including word lists, machine learning algorithms, and lexicon-based methods. However, further research is needed to enhance the accuracy and dependability of emotion detection, classification, and prediction models.

ChatGPT can provide a probabilistic measure of the affective Index and estimate the probabilistic values of Ekman's basic six human emotions in the subjective passage: happiness, sadness, anger, fear, surprise, and disgust. Comparing the similarity of an Affective Index against a list of Affective Indices through a similarity measurement creates an Affective Index Indicator list showing how similar the affective indices are between the source object and its peers. Various recommendation scenarios can apply this similarity score, such as collaborative-based, content-based, contextual-based, decision-based, and hybrid-based recommendations.

Moreover, the use of user emotion profile data raises concerns about data privacy. To address this concern, the authors suggest a separation of responsibility approach, where users safeguard their emotion profile data using an emotion ID. EARS service providers do not store user Affective Indices but instead use the user's emotion ID to generate an Affective Index Indicator list to make recommendations. However, further study is necessary to understand this scheme's operational implications and characteristics.

One can use short prompt queries to detect the presence of affective features and estimate the probabilistic values of an object's Affective Index. When combined with a separation of responsibility approach for data privacy, this holds great promise for developing more effective and ethical Emotion Aware Recommender Systems. However, further research is required to enhance the accuracy and dependability of emotion detection, classification, and prediction models. It is also necessary to evaluate the effectiveness of EARS approaches and address ethical concerns around user privacy and data protection.

## REFERENCES

Acheampong, F. A., Nunoo-Mensah, H., & Chen, W. (n.d.). Transformer models for text-based emotion detection: A review of BERT-based approaches. *Artificial Intelligence Review*, 1--41.

Bobadilla, J., Ortega, F., Hernando, A., & Gutiérrez, A. (2013). Recommender systems survey. *Knowledge-Based Systems*, *46*, 109–132.

Borth, D., Ji, R., Chen, T., Breuel, T., & Chang, S.-F. (2013). Large-scale visual sentiment ontology and detectors using adjective noun pairs. *Proceedings of the 21st ACM International Conference on Multimedia*, 223--232.

Brown, T., Mann, B., Ryder, N., Subbiah, M., Kaplan, J. D., Dhariwal, P., Neelakantan, A., Shyam, P., Sastry, G., & Askell, A. (2020). Language models are few-shot learners. *Advances in Neural Information Processing Systems*, *33*, 1877--1901.

Chang, Y.-C., & Hsing, Y.-C. (2021). Emotion-infused deep neural network for emotionally resonant conversation. *Applied Soft Computing*, *113*, 107861.

Chowdhery, A., Narang, S., Devlin, J., Bosma, M., Mishra, G., Roberts, A., Barham, P., Chung, H. W., Sutton, C., Gehrmann, S., Schuh, P., Shi, K., Tsvyashchenko, S., Maynez, J., Rao, A., Barnes, P., Tay, Y., Shazeer, N., Prabhakaran, V., … Fiedel, N. (2022). *PaLM: Scaling Language Modeling with Pathways* (arXiv:2204.02311). arXiv. http://arxiv.org/abs/2204.02311

Dale, R. (2021). GPT-3: What's it good for? *Natural Language Engineering*, *27*(1), 113--118.

Devlin, J., Chang, M.-W., Lee, K., & Toutanova, K. (2018). Bert: Pre-training of deep bidirectional

transformers for language understanding. *ArXiv Preprint ArXiv:1810.04805*.

Ethayarajh, K. (n.d.). How contextual are contextualized word representations? Comparing the geometry of BERT, ELMo, and GPT-2 embeddings. *ArXiv Preprint ArXiv:1909.00512*.

Ghazi, D., Inkpen, D., & Szpakowicz, S. (2015). Detecting emotion stimuli in emotion-bearing sentences. *Computational Linguistics and Intelligent Text Processing: 16th International Conference, CICLing 2015, Cairo, Egypt, April 14-20, 2015, Proceedings, Part II 16*, 152--165.

Grossglauser, M., Grus, J., Neumann, M., Tafjord, O., Dasigi, P., Liu, N., Hughes, M., Schmitz, M., & Zettlemoyer, L. (2018). Allennlp: A deep semantic natural language processing platform. *ArXiv Preprint ArXiv:1803.07640*.

Gujjar, J. P., & Kumar, H. P. (2021). Sentiment analysis: Textblob for decision making. *Int. J. Sci. Res. Eng. Trends*, *7*, 1097--1099.

Guo, J. (2022). Deep learning approach to text analysis for human emotion detection from big data. *Journal of Intelligent Systems*, *31*(1), 113–126. https://doi.org/10.1515/jisys-2022-0001

Hajek, P., Barushka, A., & Munk, M. (2020). Fake consumer review detection using deep neural networks integrating word embeddings and emotion mining. *Neural Computing and Applications*, *32*, 17259--17274.

Hoy, M. B. (2018). Alexa, Siri, Cortana, and more: An introduction to voice assistants. *Medical Reference Services Quarterly*, *37*(1), 81--88.

Humintell. (2020). *Ekman 7 Emotion Facial Recognition*. https://www.humintell.com/

Kanjo, E., Al-Husain, L., & Chamberlain, A. (2015). Emotions in context: Examining pervasive affective sensing systems, applications, and analyses. *Personal and Ubiquitous Computing*, *19*, 1197--1212.

Keller, J. M., Lyu, M. R., & Bourgeois, J. A. (1985). A fuzzy k-nearest neighbor algorithm. *IEEE Transactions on Systems, Man, and Cybernetics*, *4*, 580--585.

Kepuska, V., & Bohouta, G. (2018). Next-generation of virtual personal assistants (microsoft cortana, apple siri, amazon alexa and google home). *2018 IEEE 8th Annual Computing and Communication Workshop and Conference (CCWC)*, 99--103.

Kocoń, J., Cichecki, I., Kaszyca, O., Kochanek, M., Szydło, D., Baran, J., Bielaniewicz, J., Gruza, M., Janz, A., & Kanclerz, K. (2023). ChatGPT: Jack of all trades, master of none. *ArXiv Preprint ArXiv:2302.10724*.

Kompan, M., Matz, S. C., Gosling, S. D., Popov, V., & Stillwell, D. (2015). Facebook as a research tool for the social sciences: Opportunities, challenges, ethical considerations, and practical guidelines. *American Psychologist*, *70*(6), 543.

Kratzwald, B., Ilić, S., Kraus, M., Feuerriegel, S., & Prendinger, H. (2018). Deep learning for affective computing: Text-based emotion recognition in decision support. *Decision Support Systems*, *115*, 24–35.

Kusal, S., Patil, S., Choudrie, J., Kotecha, K., Mishra, S., & Abraham, A. (2022). AI-based Conversational Agents: A Scoping Review from Technologies to Future Directions. *IEEE Access*.

Kusal, S., Patil, S., Choudrie, J., Kotecha, K., Vora, D., & Pappas, I. (2022). A Review on Text-Based Emotion Detection—Techniques, Applications, Datasets, and Future Directions. *ArXiv Preprint ArXiv:2205.03235*.

Lauriola, I., Lavelli, A., & Aiolli, F. (2022). An introduction to deep learning in natural language processing: Models, techniques, and tools. *Neurocomputing*, *470*, 443--456.

Leung, J. K., Griva, I., & Kennedy, W. G. (2020a). An Affective Aware Pseudo Association Method to Connect Disjoint Users Across Multiple Datasets – an enhanced validation method for Text-based Emotion Aware Recommender. *International Journal on Natural Language Computing (IJNLC) Vol*, *9*(4). https://doi.org/10.5121/ijnlc.2020.9402

Leung, J. K., Griva, I., & Kennedy, W. G. (2020b). Making Use of Affective Features from Media Content Metadata for Better Movie Recommendation Making. *ArXiv Preprint ArXiv:2007.00636*.

Leung, J. K., Griva, I., & Kennedy, W. G. (2020c). Text-based Emotion Aware Recommender. *Proceedings of International Conference on Natural Language Computing and AI (NLCAI 2020)*, *10*, 101–114. https://doi.org/10.5121/csit.2020.101009

Leung, J. K., Griva, I., & Kennedy, W. G. (2021a). Applying the Affective Aware Pseudo Association Method to Enhance the Top-N Recommendations Distribution to Users in Group Emotion Recommender Systems. *International Journal on Natural Language Computing (IJNLC)*, *10*, 1–20. https://doi.org/10.5121/ijnlc.2021.10101

Leung, J. K., Griva, I., & Kennedy, W. G. (2021b). Making Cross-Domain Recommendations by Associating Disjoint Users and Items Through the Affective Aware Pseudo Association

Method. *Proceedings of 8th International Conference on Artificial Intelligence and Applications (AIAP 2021)*, *11*, 113–129. https://doi.org/10.5121/csit.2021.110108

Liu, Y., Ott, M., Goyal, N., Du, J., Joshi, M., Chen, D., Levy, O., Lewis, M., Zettlemoyer, L., & Stoyanov, V. (2019). Roberta: A robustly optimized bert pretraining approach. *ArXiv Preprint ArXiv:1907.11692*.

Lo, S. L., Cambria, E., Chiong, R., & Cornforth, D. (2017). Multilingual sentiment analysis: From formal to informal and scarce resource languages. *Artificial Intelligence Review*, *48*, 499--527.

Maddigan, P., & Susnjak, T. (2023). Chat2vis: Generating data visualisations via natural language using chatgpt, codex and gpt-3 large language models. *ArXiv Preprint ArXiv:2302.02094*.

Mauss, I. B., & Robinson, M. D. (2009a). Measures of emotion: A review. *Cognition and Emotion*, *23*(2), 209--237.

Mauss, I. B., & Robinson, M. D. (2009b). Measures of emotion: A review. *Cognition & Emotion*, *23*(2), 209–237. https://doi.org/10.1080/02699930802204677

Mohammad, S. M. (2016). Sentiment analysis: Detecting valence, emotions, and other affectual states from text. In *Emotion measurement* (pp. 201–237). Elsevier.

Munn, L., Magee, L., & Arora, V. (2023). Truth Machines: Synthesizing Veracity in AI Language Models. *ArXiv Preprint ArXiv:2301.12066*.

Naseem, U., Razzak, I., Musial, K., & Imran, M. (2020). Transformer based deep intelligent contextual embedding for twitter sentiment analysis. *Future Generation Computer Systems*, *113*, 58--69.

Nasir, A. F. A., Nee, E. S., Choong, C. S., Ghani, A. S. A., Majeed, A. P. A., Adam, A., & Furqan, M. (2020). *Text-based emotion prediction system using machine learning approach* (Vol. 769). IOP Publishing.

Nassiri, K., & Akhloufi, M. (2022). Transformer models used for text-based question answering systems. *Applied Intelligence*, 1--34.

Qian, Y., Zhang, Y., Ma, X., Yu, H., & Peng, L. (2019). EARS: Emotion-aware recommender system based on hybrid information fusion. *Information Fusion*, *46*, 141–146.

Russell, J. A. (2003). Core affect and the psychological construction of emotion. *Psychological Review*, *110*(1), 145.

Schedl, M., Zamani, H., Chen, C.-W., Deldjoo, Y., & Elahi, M. (2018). Current challenges and visions in music recommender systems research. *International Journal of Multimedia Information Retrieval*, *7*, 95--116.

Scherer, K. R. (2005). What are emotions? And how can they be measured? *Sage Publications Sage CA: Thousand Oaks, CA*, *44*(4), 695--729.

Song, C., Wang, X.-K., Cheng, P., Wang, J., & Li, L. (2020). SACPC: A framework based on probabilistic linguistic terms for short text sentiment analysis. *Knowledge-Based Systems*, *194*, 105572.

The design and implementation of xiaoice, an empathetic social chatbot. (2020). *MIT Press*, *46*(1), 53--93.

Thoppilan, R., Billsus, D., Hall, J., Shazeer, N., Kulshreshtha, A., Cheng, H.-T., Jin, A., Bos, T., Baker, L., & Du, Y. (2022). Lamda: Language models for dialog applications. *ArXiv Preprint ArXiv:2201.08239*.

Thorp, H. H. (2023). ChatGPT is fun, but not an author. *Science*, *379*(6630), 313--313.

Tsytsarau, M., & Palpanas, T. (2012). Survey on mining subjective data on the web. *Data Mining and Knowledge Discovery*, *24*, 478--514.

Wahde, M., & Virgolin, M. (2022). Conversational agents: Theory and applications. In *HANDBOOK ON COMPUTER LEARNING AND INTELLIGENCE: Volume 2: Deep Learning, Intelligent Control and Evolutionary Computation* (pp. 497--544). World Scientific.

Yang, K., Wang, C., Sarsenbayeva, Z., Tag, B., Dingler, T., Wadley, G., & Goncalves, J. (2021). Benchmarking commercial emotion detection systems using realistic distortions of facial image datasets. *The Visual Computer*, *37*, 1447--1466.

Yao, J. (2019). Automated sentiment analysis of text data with NLTK. *Journal of Physics: Conference Series*, *1187*(5), 052020.

Zhou, C., Li, Q., Li, C., Yu, J., Liu, Y., Wang, G., Zhang, K., Ji, C., Yan, Q., & He, L. (2023). A comprehensive survey on pretrained foundation models: A history from bert to chatgpt. *ArXiv Preprint ArXiv:2302.09419*.

Zhou, C., Li, Q., Li, C., Yu, J., Liu, Y., Wang, G., Zhang, K., Ji, C., Yan, Q., He, L., Peng, H., Li, J., Wu, J., Liu, Z., Xie, P., Xiong, C., Pei, J., Yu, P. S., & Sun, L. (2023). *A Comprehensive Survey on Pretrained Foundation Models: A History from BERT to ChatGPT*. https://doi.org/10.48550/ARXIV.2302.09419